\documentclass{INTERSPEECH2023}

\interspeechcameraready

\title{Gender-ambiguous voice generation through feminine speaking style transfer in male voices}
\name{Maria Koutsogiannaki$^1$, Shafel Mc Dowall$^1$, Ioannis Agiomyrgiannakis$^2$}
\address{
  $^1$Crayon Data \& AI Center of Excellence\\
  $^2$Altered.ai }
\email{maria.koutsogiannaki@crayon.com, shafel.mcdowall@crayon.com, agios@altered.ai}

\usepackage{multirow}

\begin{document}

\maketitle 
\vspace{-0.2in}
\begin{abstract}
	\vspace{-0.05in}
	Recently, and under the umbrella of Responsible AI, efforts have been made to develop gender-ambiguous synthetic speech to represent with a single voice all individuals in the gender spectrum. However, research efforts have completely overlooked the speaking style despite differences found among binary and non-binary populations. In this work, we synthesize gender-ambiguous speech by combining the timbre of a male speaker with the manner of speech of a female speaker using voice morphing and pitch shifting towards the male-female boundary. Subjective evaluations indicate that the ambiguity of the morphed samples that convey the female speech style is higher than those that undergo plain pitch transformations suggesting that the speaking style can be a contributing factor in creating gender-ambiguous speech.  To our knowledge, this is the first study that explicitly uses the transfer of the speaking style to create gender-ambiguous voices.
\end{abstract}

\vspace{-0.05in}
\section{Introduction}
\vspace{-0.05in}
Traditionally, voice assistants have been designed with a female voice. However, this has led to a reinforcement of gender stereotypes and a lack of inclusiveness for those who identify themselves as non-binary. 

Lately, there have been initiatives within the Responsible AI framework to create a synthetic voice that is gender-ambiguous and can represent all individuals across the gender spectrum. Perhaps, the most well-know effort is the release of the genderless voice assistant Q \cite{meetQ}. And while the introduction of Q highlighted the need for gender-neutral voices, it did not resolve any gender-related Responsible AI issues. In fact, it revealed gaps in the definition, implementation, and evaluation of what we refer to as gender-ambiguous voice \cite{sutton2020gender}. 

 In this work, we try to address the aforementioned gaps. First, we analyse past research on speech differences among male, female, and non-binary speakers and identify pitch, timbre, and speaking style as the primary factors that influence gender perception. Then, we note that none of the existing research considers speaking style as a factor in generating gender-ambiguous voices. Thus, we introduce a novel approach to create an ambiguous synthetic voice that takes into consideration both acoustic and suprasegmental characteristics. Specifically, we perform voice morphing of a female voice to a male voice. The morphed voice combines the feminine speaking style with the male timbre while the pitch of the morphed voice is modified to fall within the pitch boundary of male-female voices. Equally important, is the assessment of the candidate gender-ambiguous voices. We establish a clear definition of what is gender-ambiguous, as this is currently lacking in the existing literature, to be able to define the metrics for the assessment. Last, we design an evaluation framework to avoid bias during the assessment process. Results show that our proposed method effectively produces gender-ambiguous speech. The generated samples are of good quality and require 20 min of speech audio from the target speaker which makes our voice conversion methodology appealing as a post-process of Text-To-Speech (TTS) synthesis systems. To our knowledge, this is the first study that explicitly uses the speaking style as a key factor in creating gender-ambiguous voices, provides a clear definition of what is gender-ambiguous voice and sets the responsible AI framework for its assessment. 

\vspace{-0.05in}
\section{Related work}
\vspace{-0.08in}
\subsection{Speech differences among binary and non-binary}
\vspace{-0.08in}
There is a growing number of individuals (genderqueer, genderfluid, agender, bigender, etc.) who do not identify themselves within the traditional male-female gender binary. These non-binary individuals do not conform to traditional gender norms or expectations, including the way they dress, the way they behave, but also the way they speak.  Non-binary individuals may modify their speaking style in order to better align with their gender identity by adjusting their pitch, tone, and vocal inflection to create a more gender-neutral speaking style. Thus, the term "gender-ambiguous" did not arise solely from the need of a single voice to represent all genders in voice systems but is an inherent property of the voice of many non-binary individuals.

Various studies have underlined speech differences between binary and non-binary populations. Acoustically, there have been found differences in mean fundamental frequency F0, significantly lower than the cis women's and significantly higher than the cis men's, significantly less negative spectral slope and higher cepstral peak prominence-smoothed, indicative of a brighter, more resonant voice quality \cite{brown2022}. The study of \cite{Brandon2023} also found that fundamental frequency F0 and formant centroids of non-binary individuals fell within an intermediate range of cisgender men and women while other measures (e.g. cepstral peak prominence) demonstrated more complex patterns in relation to the other speaker groups (transgender men).

Differences in suprasegmental characteristics were also found. The intonation patterns of non-binary populations did not pattern like women nor like men but they patterned with a combination of both, indicating either a mix of feminine and masculine traits or a rejection of both \cite{maxwell2019}. In \cite{Samuelsson2007GenderEO}, non-binary linguistic patterns have been found to combine masculine (smaller pitch range) and feminine features (high rise terminals)  and this combination made these individuals sound gender-ambiguous.

\vspace{-0.05in}
\subsection{Gender-ambiguous voice synthesis}
\vspace{-0.1in}
The combination of feminine and masculine patterns seems to be the key to synthesize a gender-ambiguous voice. However, the idea of mimicking how non-binary people elicit gender-ambiguous speech has not yet been explored, despite the fact that previous studies have shown the benefits of speech modifications techniques that are inspired by human speaking style adaptation to different communication scenarios \cite{koutsogiannaki15_interspeech, godoy}.


 In the field of gender-ambiguous voice synthesis, we found two main research lines. The first line focuses on modifying the specific acoustic features of the speech signal to make it sound gender-ambiguous. In \cite{electronics11101594}, pitch and formant modifications are performed on the original signal using fixed step sizes within a range of frequencies. The optimal pitch and formant shift parameters are selected by a speaker gender recognition model that outputs the certainty for each classification. 

The second research line synthesizes speech from text by averaging model embeddings from male and female speakers \cite{Yu2022, Stoidis2022,  raftis2023}. In \cite{raftis2023} authors generate gender-ambiguous voices by sampling on the speaker embedding space of a multilingual multi-speaker non-attentive Tacotron model. They used more than 1000 hours of audio and equivalent amount of binary speakers, balanced as regards the speaker gender within each language to train the model. Results showed higher gender-ambiguity compared to baseline.

However, there is a third research line indirectly linked to the creation of gender-ambiguous voice, proposed back in 2012 \cite{pernet2012}. The intention of authors was not to create gender-ambiguous speech for TTS systems but to understand which feature contributes more to the gender perception, pitch or timbre. Authors used syllables of male end female speakers and they morphed them using STRAIGHT to create a  male/female continuum where the pitch or timbre was held constant. According to this article, voice gender categorization does not depend on pitch perception but can be performed using timbre information only and pitch is used only when timber information is ambiguous. \cite{pernet2012}.
\vspace{-0.05in}
\subsection{Motivation}
\vspace{-0.05in}
Reviewing the literature on speech differences within and between binary and non-binary populations and the existing work on gender-ambiguous voice generation we come to the following conclusions that motivated us to propose this work. 

First, implementing some of the strategies non-binary people adopt to create gender-ambiguity in their speech could be proven also beneficial in the generation of gender-ambiguous voices. One such strategy seems to be the combination of feminine and masculine patterns. 

Second, merely acoustic modifications to the speech signal as proposed in \cite{electronics11101594} are not sufficient to create gender-ambiguous speech.  There are suprasegmental differences (speaking rate, intonation patterns, pause distribution, pitch contour etc.)  between binary populations and "abudant gender signalling information in the speech" \cite{sutton2020gender} important for gender identification.

Third, averaging model embeddings from male and female speakers could be a better approach to create ambiguous speech since indirectly suprasegmental characteristics of both genders could be also averaged. However, the main drawback in the process is data acquisition, that is many hours of many male and female voices \cite{raftis2023}. This requires a lot of effort especially if we consider that each time we need to perform an adaptation (e.g. language adaptation) the dataset needs to be rebuilt and the model to be retrained.

Voice conversion appears to be the best approach for tackling the aforementioned limitations and achieving gender-ambiguous synthetic voices. Going beyond gender ambiguity on syllables \cite{pernet2012} which clearly does not reveal much of the speaking style, we believe that the blending of feminine speaking style with masculine timbre with some modifications of the male pitch voice towards the male-female overlapping pitch boundary could create gender ambiguity in the voice. The advantage of the method is that can be applied as a post-processing technique to any TTS system developed. Moreover, modeling the target speaker requires 20 min of speech samples and unlike other techniques that average the embedding space \cite{raftis2023} there is more control over the timbre characteristics of the voice created.

\vspace{-0.05in}
\section{Methodology}
\vspace{-0.05in}
\subsection{Gender-ambiguous voice generation}
\vspace{-0.05in}
To generate gender-ambiguous voices, we used the Bella synthetic voice from Microsoft Azure TTS service to create female synthetic speech samples from transportation announcements (to avoid gender bias in the text content, such as he/she pronouns and emotional text). The synthesized samples had an average duration of 11 seconds each. We then morphed Bella's samples to two male target voices, Taylor (pitch 140Hz) and James (pitch 132Hz) models using our voice morphing tool, Altered Studio \cite{alteredai}. The resulting samples were further shifted 3 and 4 semitones respectively towards the male-female boundary (170Hz) to increase ambiguity. Two gender-ambiguous candidate voices resulted from this process,  mTaylor and mJames.

Our hypothesis is that the ambiguity of the candidate voices, mTaylor and mJames, does not only come from the pitch modification, but also from the transfer of the female speaking style. Thus, we need to examine the effect of pitch shifting in isolation on the gender-ambiguity of the samples. Therefore, we create pitch-shifted samples of Taylor and James (pTaylor, pJames) as follows: using the altered.ai TTS system, we synthesize the 5 transportation audio samples with the voice of Taylor and James (Taylor, James) and pitch-shift the samples (pTaylor, pJames) using the same algorithm and pitch-shifting values (3 and 4 semitones respectively).

\vspace{-0.05in}
\subsection{Evaluation framework}
\vspace{-0.05in}
To evaluate the ambiguity of the produced samples and our main hypothesis, that the transfer of the female speaking style in male voices creates higher gender-ambiguity compared to pitch modifications alone, we designed an evaluation framework consisted of 3 listening tests to avoid bias in the process (Table~\ref{table:agents}).

Listening test A evaluates our gender-ambiguous candidate voices, mJames and mTaylor, produced with our proposed method described above. Listeners evaluate the 5 transportation announcements spoken by mJames and mTaylor and also by 2 male and 2 female voices (a total of 30 samples: 6 speakers x 5 sentences). The male (Alfie, Ryan) and female (Abbi, Sonia) voices were synthesized using Azure TTS and their presence in the listening test serves to establish a baseline perception of gender and make statistical comparisons with the morphed samples. We did not include Bella's voice to avoid bias (speaking style transfer) in the evaluation of mJames and mTaylor.

Listening test B used the same stimuli as A, but here the  pitch-modified voices, pJames and pTaylor, replace the morphed samples, mJames and mTaylor. The test was conducted with the same text sentences and presented to listeners at least 5 days after the conduction of listening test A to reset their perception (exclude memory bias). We did not present p-voices with m-voices in the same test to avoid listeners assigning scores based on their familiarity with the voice, rather than the features being evaluated.

Listening test C includes the original source (Bella) and target speakers (James and Taylor). This is a control test to ensure that the voices we used for voice morphing are indeed binary and not gender-ambiguous. Again here, we include the binary voices (Abbie, Sonia, Alfie, Ryan) to cover the gender spectrum. To keep the test duration shorter, we randomly selected 1 transportation announcement out of the 5 for each voice and this randomization was performed per listener.

\begin{table}[]
	\centering
	\fontsize{7}{9}\selectfont
	\setlength{\tabcolsep}{4pt}
	\renewcommand{\arraystretch}{1}
	\caption{Overview of the synthetic voices involved in the listening tests and the methods used to create them.}
	\label{table:agents}
	\begin{tabular}{|c|c|c|c|}
		\hline
		Method                                                                                 & Agent   & Description                                                                                           & \begin{tabular}[c]{@{}c@{}}Listening \\ test\end{tabular} \\ \hline
		\multirow{5}{*}{Azure TTS}                                                             & Abbie   & Female                                                                                                & A,B                                                       \\ \cline{2-4} 
		& Sonia   & Female                                                                                                & A,B                                                       \\ \cline{2-4} 
		& Alfie   & Male                                                                                                  & A,B                                                       \\ \cline{2-4} 
		& Ryan    & Male                                                                                                  & A,B                                                       \\ \cline{2-4} 
		& Bella   & Female                                                                                                & C                                                         \\ \hline
		\multirow{2}{*}{\begin{tabular}[c]{@{}c@{}}Altered.ai\\ TTS\end{tabular}}              & James   & Male                                                                                                  & C                                                         \\ \cline{2-4} 
		& Taylor  & Male                                                                                                  & C                                                         \\ \hline
		\multirow{4}{*}{\begin{tabular}[c]{@{}c@{}}Altered.ai\\ Voice conversion\end{tabular}} & mJames  & \begin{tabular}[c]{@{}c@{}}source speaker Bella\\ target speaker James\\ pitch shifting 4 semitones\end{tabular}  & A                                                         \\ \cline{2-4} 
		& mTaylor & \begin{tabular}[c]{@{}c@{}}source speaker Bella\\ target speaker Taylor\\ pitch shifting 3 semitones\end{tabular} & A                                                         \\ \cline{2-4} 
		& pJames  & \begin{tabular}[c]{@{}c@{}}source speaker James\\ pitch shifting 4 semitones\end{tabular}                         & B                                                         \\ \cline{2-4} 
		& pTaylor & \begin{tabular}[c]{@{}c@{}}source speaker Taylor\\ pitch shifting 3 semitones\end{tabular}                        & B                                                         \\ \hline
	\end{tabular}
	\vspace{-0.2in}
\end{table}

\vspace{-0.1in}
\subsubsection{Evaluation metrics}
Equally important to the presentation of the stimuli are the metrics used to evaluate the gender-ambiguity in the voice. However, the lack of definition of what is gender-ambiguous makes it difficult to determine them. So what is a gender-ambiguous voice? 
To start with, we believe that is highly improbable that the listener will question the gender of the voice that listens to unless explicitly asked to do so. People tend to automatically map a gender to the voice they listen to  \cite{sutton2020gender} and the gender selected depends on the  listener's cognitive perception and life experience \cite{keith2006}.  Considering the above, we provide our definition of what is a gender-ambiguous voice:
\underline{Definition 1}:	Gender-ambiguous voice can be considered the voice that cannot be classified with certainty as male or female by one listener when the listener is asked.
\underline{Definition 2}: Gender-ambiguous voice can be considered the voice that has equal probability of being classified as male or female across many listeners.

Based on definition 2, we prompt participants to classify gender as either male or female and we anticipate an equal distribution for gender-ambiguous voices across listeners. Based on definition 1, we ask participants to rate their confidence in their classification and level of surprise in case of miss-classification. Additionally, we include questions regarding the perceived femininity and masculinity of the voice to understand how these factors influence decision-making. Finally, we ask participants to rate the quality of the synthesized samples since our aim is to use our proposed method for production environments. Table \ref{table:metrics} contains the metrics presented to the listeners. 

\begin{table}[htbp]
	\centering
	\fontsize{7}{9}\selectfont
	\setlength{\tabcolsep}{1pt}
	\renewcommand{\arraystretch}{1}
	\caption{Evaluation metrics presented to the listeners}
	\label{table:metrics}
	\vspace{-0.1in}
	\begin{tabular}{|c|p{8cm}|p{35mm}|}
		\hline
		ID & Metric and Score \\ \hline
		Q1 & How would you identify the person speaking? Male/Female  \newline male: -1, female: 1 \\ \hline
		Q2 & How confident are you for your answer?  \newline Very confident: 5 , Confident: 4 , Somewhat confident: 3 , Not so confident: 2 , Not confident at all: 1 \\ \hline
		Q3 & if you identify the person speaking as male/female, how surprised would you be if it was actually female/male? \newline Very surprised: 4, Surprised: 3, Not so much surprised: 2, Not at all surprised: 1 \\ \hline
		Q4 & Rate the femininity and masculinity of the voice \newline Very masculine: -3, Masculine: -2, Fair masculine: -1, Neither masculine nor feminine: 0, Fair feminine: 1, Feminine: 2, Very feminine: 3 \\ \hline
		Q3 & Rate the overall quality of the sound \newline Excellent: 5, Good: 4, Fair: 3, Poor: 2, Bad: 1\\ \hline
	\end{tabular}
	\vspace{-0.2in}
\end{table}

\vspace{-0.1in}
\section{Evaluation}
Subjective evaluations of gender-ambiguous candidate voices were conducted by 35 listeners of different binary gender (14 female, 21 male), nationality (16 nationalities) and linguistic background (3 languages on average) to ensure a variability in the cognitive perception and life experience. All of them had a good level of English language. Unlike the study of \cite{raftis2023} where the gender-ambiguous samples were evaluated by native speakers, in this study we use English as language and we ask participants of different nationalities to provide their gender perception. The decision for this is that many TTS systems address to a public of different nationalities in English. So in this work, we examine the gender perception mainly from the aspect of non-nativeness. The samples as stated before were balanced, of equal number of male, female, and gender-ambiguous candidate samples, presented in a controlled (not 3 samples from same speaker presented sequentially) random order across evaluators. Participants are asked to perform first the listening test A\footnote{To listen to our gender-ambiguous candidate voices you may visit \url{www.csd.uoc.gr/~mkoutsog/gender_ambiguous.html}}, then the listening test C and after 5 days the listening test B. It could be the case that we introduced some bias to the listening test C (the voices of Taylor and James could be considered more feminine after listening to the voices of mTaylor and mJames in the listening test A), but we decided to go with this design to keep our participants engaged since we estimated it would not insert any bias in our principal test A, neither in the test B which was conduced with delay for this reason.  

\begin{figure}[h]
	\vspace{-0.1in}
	\centering
	\includegraphics[width=\linewidth]{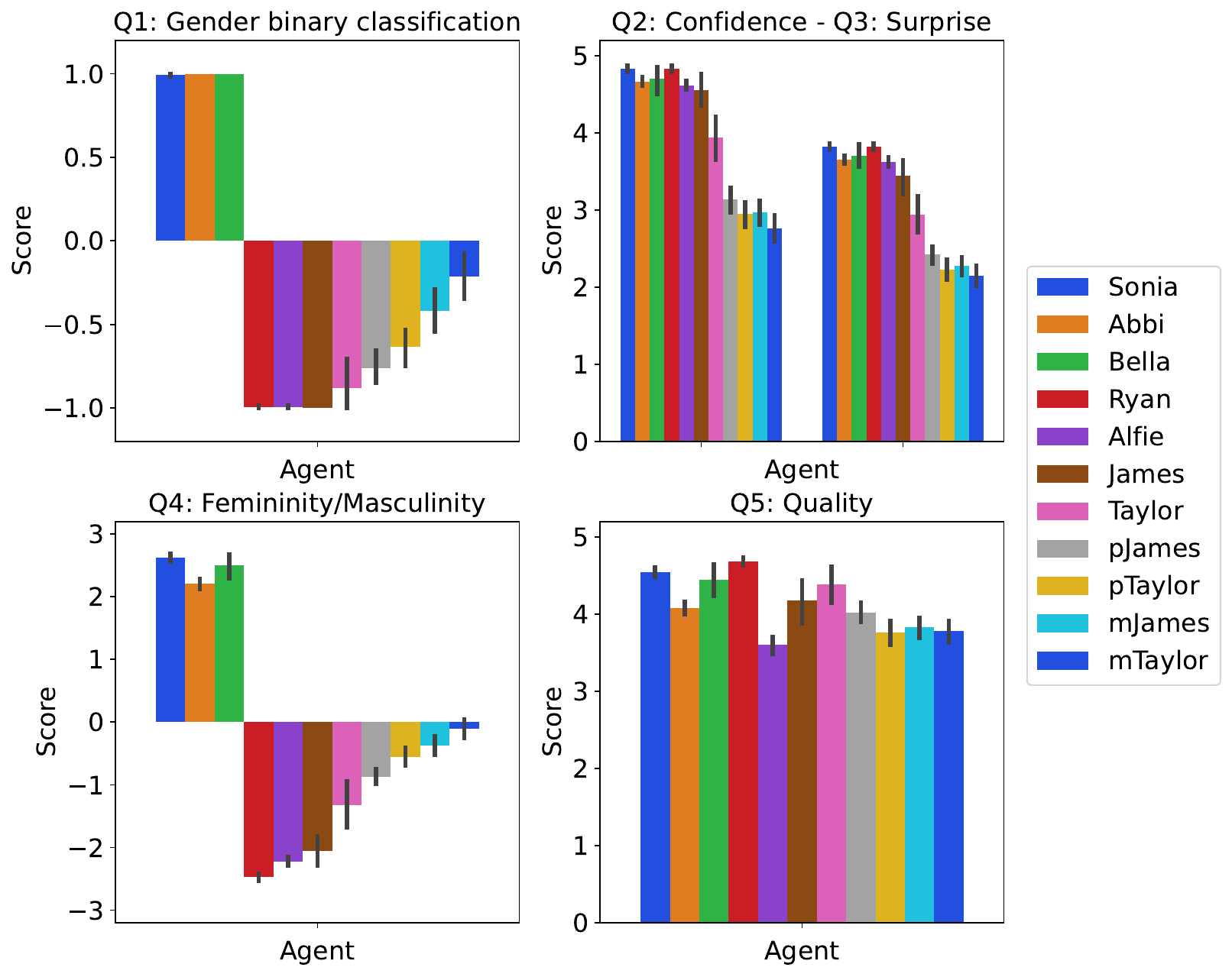}
	\caption{Subjective evaluation scores one each metric Q1: gender classification, Q2: Confidence, Q3: Surprise, Q4: Femininity/masculinity, Q5: Quality as described in Table~\ref{table:metrics} of all 3 listening tests A, B and C. The proposed gender-ambiguous voices are mTaylor and mJames which resulted from the voice morphing of Bella (source speaker) and Taylor and James (target speakers) respectively. pTaylor and pJames derived by pitch-shifting 3 and 4 semitones the voices of Taylor and James respectively. Ryan, Alfie, Sonia and Abbi are binary voices used to balance the listening test and for statistical analysis.}
	\label{fig:evaluation}
	\vspace{-0.2in}
\end{figure}

\begin{table}[h]
	\hspace{-5pt}
	\fontsize{7}{9}\selectfont
	\setlength{\tabcolsep}{1pt}
	\renewcommand{\arraystretch}{1.1}
	\caption{Posthoc test analysis  on each metric among different agents. Significance $p<0.0001$ is indicated with $****$, $p<0.001$: $***$, $p<0.01$: $**$, $p<0.05$: $*$}
	\begin{tabular}{|l|l|l|l|l|l|}
		\hline
		\multicolumn{1}{|c|}{Agent} & \multicolumn{1}{c|}{Abbi / Sonia}                                             & \multicolumn{1}{c|}{Ryan / Alfie}                                                                           & \multicolumn{1}{c|}{mJames}                                        & \multicolumn{1}{c|}{mTaylor}                                                                          & \multicolumn{1}{c|}{pJames} \\ \hline
		mJames                      & \begin{tabular}[c]{@{}l@{}}Q1-Q4:  ****\\ Q5:        p=0.27/****\end{tabular} & \begin{tabular}[c]{@{}l@{}}Q1-Q4:  ****\\ Q5:        ****/p=1\end{tabular}                                  &                                                                    &                                                                                                       &                             \\ \hline
		mTaylor                     & \begin{tabular}[c]{@{}l@{}}Q1-Q4:  ****\\ Q5:        p=0.09/****\end{tabular} & \begin{tabular}[c]{@{}l@{}}Q1-Q4:  ****\\ Q5:        ****/p=1\end{tabular}                                  & Q1-Q5:  p=1                                                        &                                                                                                       &                             \\ \hline
		pJames                      & \begin{tabular}[c]{@{}l@{}}Q1-Q4:  ****\\ Q5:        p=1/****\end{tabular}    & \begin{tabular}[c]{@{}l@{}}Q1:        p\textgreater{}0.253\\ Q2-Q4:  ****\\ Q5:        ****/**\end{tabular} & \begin{tabular}[c]{@{}l@{}}Q1:        *\\ Q2-Q5:  p=1\end{tabular} & \begin{tabular}[c]{@{}l@{}}Q1:        ****\\ Q2-Q3:  p=1\\ Q4:        *\\ Q5:        p=1\end{tabular} &                             \\ \hline
		pTaylor                     & \begin{tabular}[c]{@{}l@{}}Q1-Q4:  ****\\ Q5:        p=0.16/****\end{tabular} & \begin{tabular}[c]{@{}l@{}}Q1:        *\\ Q2-Q4:  ****\\ Q5:        ****/p=1\end{tabular}                   & Q1-Q5:  p=1                                                        & \begin{tabular}[c]{@{}l@{}}Q1:         **\\ Q2-Q5: p=1\end{tabular}                                    & Q1-Q5: p=1                  \\ \hline
	\end{tabular}
	\label{table:posthoc}
	\vspace{-0.1in}
\end{table}

Figure \ref{fig:evaluation} summarises the results on the 5 questions (Table \ref{table:metrics}) presented to the participants.   
A Kruskal-Wallis test was performed to compare the scores between the eight groups \{Alfie, Ryan, Abbi, Sonia, mJames, mTaylor, pJames, pTaylor\}. Test C results were excluded from the statistical analysis due to lower sample size compared to A and B. Besides test C was used as a control to ensure that the source and target voices were indeed binary. The statistical test analysis indicated a significant difference in the scores for the gender identification (question Q1, $H(7) = 1774$, $p < 0.0001$), the level of confidence  (question Q2, $H(7) = 1186$, $p < 0.0001$), the level of surprise (question Q3, $H(7) = 1160$, $p < 0.0001$), the rate of femininity-masculinity (question Q4, $H(7) = 1724$, $p < 0.0001$) and the quality (question Q5, $H(7) = 377$). 

To further investigate the differences between groups, a Dunn's post-hoc test was performed with Bonferroni correction (See Table \ref{table:posthoc}). Pairwise comparisons showed significant differences across binary groups ($p<0.0001$, e.g. Ryan and Sonia) and no differences within the same binary group ($p=1$, e.g Ryan and Alfie) on the questions related to gender identification (Q1) and femininity/masculinity (Q4). 

Our primary analysis focused on the candidate gender-ambiguous voices. Results on gender classification (Q1) showed that mJames had significantly different scores from both male and female agents ($p<0.0001$), whereas pJames had significantly different scores only from female agents ($p<0.0001$) but not from male agents ($p>0.253$). This finding  suggests that the pitch transformation used on James did not effectively differentiate him from other males. However, incorporating feminine speaking style with his masculine traits resulted in greater ambiguity across listeners. 

Similarly to mJames, mTaylor's scores were found significantly different (Table \ref{table:posthoc}) from binary agents ($p<0.0001$) both male and female on Q1 suggesting also that mTaylor's voice is gender-ambiguous. Also, pTaylor had scores significantly different from the females ($p<0.001$) and males ($p<0.05$). Comparing mTaylor and pTaylor's scores, it can be seen that mTaylor's voice introduces a greater ambiguity compared to pTaylor and this result is significant ($p<0.01$). The increased level of ambiguity resulting from pitch modifications in Taylor's voice and the even greater level of ambiguity resulting from morph modifications can be attributed to the fact that Taylor has relatively weaker male characteristics compared to mJames (higher pitch) and possibly timbre. This is also supported by the fact that Taylor's voice in the post listening test (C) was perceived by some listeners as fair masculine and also some few listeners classified him as female in the gender classification task (even though classification to female class is not supported statistically) and with less masculine traits in the Q4 task (significant difference is found with the female agents and with Ryan). Despite the fact that there can be some post effects of listening test A to listening test C in the perception of the masculinity of Taylor's and Jame's voice, we do believe that strong male voices could be more challenging in creating balanced ambiguity. For example, low pitch voices would require greater shift adjustments towards the ambiguous pitch area of 170Hz and higher pitch shifts would result to lower voice quality. Therefore, the selection of the target speaker is important for the quality of the outcome. On average, our proposed method creates a 0.6 and 0.3 drop on the quality of the voices of Taylor and James respectively and produces samples of good quality ($(m, sd)=(3.8, 0.9)$)  for both mJames and mTaylor. 


An interesting finding is that metacognition tasks involving certainty and surprise make no distinction between morphed and pitched voices. The listeners' certainty is affected both by the transformations that combine the feminine with the masculine traits and by merely acoustic transformations. Moreover,  a lower confidence score can also be seen for binary agents. This implies that listeners' confidence in determining gender is influenced not only by voice characteristics but also by other metacognition processes (self confidence etc.).  Reporting lower confidence may be an easier task than being asked to make a decision. This suggests, that gender-ambiguous assessment should not include merely certainty as metric but should be assessed by the listeners' variability on the gender classification task. This is of great importance since the majority of the studies use a certain to probable male/female scale. Another interesting finding is that the metric of femininity and masculinity highly correlates with that of the gender classification task. 

Last, we examined for significant differences in the responses between male and female participants. The Mann-Whitney U test was conducted separately for each question after dividing the data into two groups based on the gender of the participants. Test analysis revealed a significant participant's gender effect on scores for the agent pJames ($U = 2631$, $p < 0.01$), mTaylor ($U=2979$, $p<0.05$) and Abbi ($U=15805$, $p<0.05$) on Q2 question, for the agents Alfie ($U = 16497$, $p < 0.01$) and mTaylor (U = 3042.00, $p < 0.05$) on question Q3, for Ryan ($U=17725$, $p<0.0001$) and Sonia ($U=12715$, $p<0.05$) on question Q4 and for Ryan ($U=12445$, $p<0.05$) on question Q5. This suggests that male and female participants may differ in their metacognition state and this state could be influenced by the characteristics of the agent e.g for question Q4 it could be differently how women perceive masculinity and femininity compared to men.

\vspace{-0.05in}
\section{Conclusion}
\vspace{-0.05in}
This study presents a novel approach of generating gender-ambiguous speech by blending feminine and masculine voice traits. We used voice morphing to transfer feminine speaking style to male voices which resulted in higher levels of ambiguity than using pitch transformations alone. We developed a framework of how to assess gender-ambiguous candidate voices without adding bias in the process and we show that metacognition metrics of certainty should not be used alone but in combination with a gender classification task across many participants.  
This research highlights the significance of speaking style in producing gender-ambiguous voices and sets a foundation for further investigation beyond the transfer of suprasegmental features from feminine to masculine.

%

%
%

\bibliographystyle{IEEEtran}
\bibliography{mybib}

\end{document}